\documentclass[preprint,showpacs,preprintnumbers,amsmath,amssymb]{revtex4}
\usepackage{graphicx}
\usepackage{dcolumn}
\usepackage{bm}
\usepackage{epstopdf}%

\begin{document}

\title{Graphene on Rh(111): combined DFT, STM, and NC-AFM studies}

\author{Petar Stojanov,$^1$ Elena Voloshina,$^2$ Yuriy Dedkov,$^1$ Stefan Schmitt,$^1$ Torben Haenke,$^1$ and Andreas Thissen$^1$}
\affiliation{$^1$SPECS Surface Nano Analysis GmbH, Voltastra\ss e 5, 13355 Berlin, Germany}
\affiliation{$^2$Physikalische und Theoretische Chemie, Freie Universit\"at Berlin, 14195 Berlin, Germany}

\begin{abstract}
The simultaneous combination of scanning probe methods (tunnelling and force microscopies, STM and AFM) is a unique way to get an information about crystallographic and electronic structure of the studied surface. Here we apply these methods accompanied by the state-of-the-art density functional theory (DFT) calculations to shed a light on the structure and electronic properties of the strongly-corrugated graphene/Rh(111) system. The atomically resolved images are obtained for both STM and AFM modes and compared with the DFT results showing good agreement between theory and experiment.
\end{abstract}

\keywords{atomic force microscopy; density functional theory calculations; graphene; moir\'e structures; Rh(111); scanning tunnelling microscopy}


\maketitle

\section{Introduction}

The discovery of the fascinating properties of graphene~\cite{Geim:2011, Novoselov:2011} initiates the fountain of experimental and theoretical works devoted to studies of its electronic and structural properties. In most cases a graphene layer in these works is in contact with the supporting substrate: insulating, semiconducting or metallic. The latter supports are considered as the most interesting from technological and fundamental-science points of view and several recent reviews have pointed out the importance of these graphene/metal systems~\cite{Wintterlin:2009,Batzill:2012,Dedkov:2012book,Voloshina:2012c}. Firstly, the preparation of graphene on metals is presently considered as a cheapest way to produce large graphene layers of different thickness that can be transferred onto the polymer or insulating support and then used for preparation of different devices~\cite{Bae:2010,Lin:2010,Lin:2011}. Another interesting, more fundamental, aspect of the graphene-metal interface is the nature of interaction at the interface~\cite{Voloshina:2012c}. Here, the intuitive description divides the whole set of the graphene-metal interfaces on two large subclasses of the \textit{weakly} and \textit{strongly} interacting graphene with metal. As a measure of interaction strength one can consider the bonding energy per carbon atom between graphene and metal, the doping level of graphene, or/and the state and position of the Dirac cone in the electronic structure of the system. Interesting to note that this ``graphene-metal interaction'' puzzle is valid for a graphene layer adsorbed on metal as well as for the metal layers deposited on graphene (free-standing or substrate-supported).

Among the graphene/metal interfaces, the graphene layers on the lattice-mismatched close-packed surfaces [Rh(111), Ru(0001), Ir(111), Pt(111)] are under intent attention~\cite{Wang:2010ky,Sicot:2010,Voloshina:2012a,Marchini:2007,Coraux:2009,Busse:2011,Sutter:2009a,Land:1992}. The interaction between graphene and metal is abruptly changed from strong to weak interaction when going from Rh or Ru to Ir or Pt. Moreover, these systems are interesting from the perspective that metal atoms deposited on top of such graphene/metal systems will form an ordered arrays of clusters due to the moir\'e structure of the graphene layer on the lattice-mismatched surfaces. In such moir\'e structures the geometry of the system (buckling of a graphene layer) as well as the space variation of the charge (potential) will define the structure and the electronic properties of the metal-clusters/graphene/metal systems.

Here we performed the studies of the crystallographic and electronic structure of the strongly-corrugated graphene layer on Rh(111) via combination of scanning probe techniques (STM and AFM) and density-functional theory (DFT) calculations. These results show that a graphene layer in this system is strongly corrugated and electron density is strongly localised at the graphene/Rh(111) interface in places where graphene is strongly bonded to the substrate. Our results give also an information on the role of dispersive forces (van der Waals) in the bonding mechanism in this system and help to understand the clustering mechanism on top of strongly corrugated graphene layer.

\section{Experimental}

The electronic and structural properties of the graphene/Rh(111) system are obtained using generalised gradient approximation as parameterized by Perdew~\textit{et al.} (PBE)~\cite{Perdew:1996} to the exchange correlation potential. For solving the resulting Kohn-Sham equation we use the Vienna Ab Initio Simulation Package (VASP)~\cite{Kresse:1994} with the projector augmented wave basis sets~\cite{Blochl:1994}. The long-range van der Waals interactions were accounted for by means of a DFT-D2 approach proposed by Grimme~\cite{Grimme:2004,Grimme:2006,Grimme:2010,Dion:2004}. This method relies on corrections added to the DFT total energy and forces, based on a damped atom-pairwise potential $C_6R^{-6}$ ($C_6$ represents the dispersion coefficient for a given atom pair and $R$ is the distance between the atoms). The studied system is modelled using supercell, which has an $(11\times11)$ lateral periodicity and contains one layer of $(12\times12)$ graphene on four-layer slab of metal atoms. Metallic slab replicas are separated by ca. 18\,\AA\ in the surface normal direction, leading to an effective vacuum region of about 15\,\AA. To avoid interactions between periodic images of the slab, a dipole correction is applied~\cite{Neugebauer:1992}. Due to the large lateral periodicity the surface Brillouin zone is sampled with a single $k$-point at $\Gamma$ for structure optimisation (the positions of the carbon atoms as well as those of the top three layers of Rh are optimised) and set to $3\times3\times1$ mesh in the total energy calculations. The Rh-Rh spacings in the bottom layer are fixed at the optimised bulk value. The STM images are calculated using the Tersoff-Hamann formalism~\cite{Tersoff:1985}, which states that the tunnelling current in an STM experiment is proportional to the local density of states (LDOS) integrated from the Fermi level to the bias. In the visualisation software Hive~\cite{Vanpoucke:2008} it is implemented in its most basic formulation, approximating the STM tip by an infinitely small point source. The integrated LDOS is calculated as $\bar{\rho}(\mathbf{r},\varepsilon)\propto\int_\varepsilon^{E_F}\rho(\mathbf{r},\varepsilon')d\varepsilon'$
with $E_F$ the Fermi energy. An STM in constant current $\varepsilon$ mode follows a surface of constant current, which translates into a surface of constant integrated [$\bar{\rho}(x,y,z,\varepsilon)=C$ with $C$ a real constant]. For each $C$, this construction returns a height $z$ as a function of the position $(x, y)$. This height map is then mapped linearly onto a corresponding colour scale.
In AFM images simulations, the tip-sample force is expressed as a function of the potential $V_{ts}\,(\mathbf{r})$ on the tip due to the sample: $F_{ts}\propto - \nabla[|\nabla V_{ts}(\mathbf{r})|^2]$ and transferred to the frequency shift of the oscillating sensor via formalism described in Ref.~\cite{Chan:2009jb}.

The graphene/Rh(111) system was prepared in ultra-high vacuum station for STM/AFM studies via cracking of propylene gas (C$_3$H$_6$) according to the recipe described in details in Refs.~\cite{Sicot:2010,Sicot:2012,Voloshina:2012a}. Prior to the graphene preparation the Rh(111) was cleaned via cycles of Ar$^+$-sputtering and high-temperature annealing until the clean atomically-resolved images were acquired in STM. Further, the quality and homogeneity of the graphene/Rh(111) samples were verified by means of low-energy electron diffraction (LEED) and STM. The STM/AFM images were collected with Aarhus SPM 150 equipped with KolibriSensor\texttrademark\
from SPECS~\cite{specs,Torbruegge:2010cf} with Nanonis Control system. In all measurements the sharp W-tip was used which was cleaned \textit{in situ} via Ar$^+$-sputtering. In presented STM images the tunnelling bias voltage, $U_T$, is referenced to the sample and the tunnelling current, $I_T$, is collected by the tip, which is virtually grounded. During the AFM measurements the sensor was oscillating with the resonance frequency of $f_0 = 1001541$\,Hz and the quality factor of $Q = 32323$, and the frequency shift was used as an input signal in a feedback loop for the topography
measurements. The oscillation amplitude was set to $A = 300$\,pm. The system base pressure was better than $8\times10^{-11}$\,mbar during all experiments. All measurements were performed at room temperature.

\section{Results and Discussion}

\begin{figure}
\caption{(a) Ball model of the graphene/Rh(111) system. Circles, triangles, rectangles, and starts denote the high-symmetry positions for carbon atoms on Rh(111). (b,c) Top and side views, respectively, of the difference electron density maps, $\Delta n(r) = n_{gr/Rh}(r) - n_{Rh}(r) - n_{gr}(r)$, ploted in units of $e/$\AA$^3$ calculated for graphene/Rh(111).}
\label{grRh_struct}
\includegraphics[width=9cm,keepaspectratio]{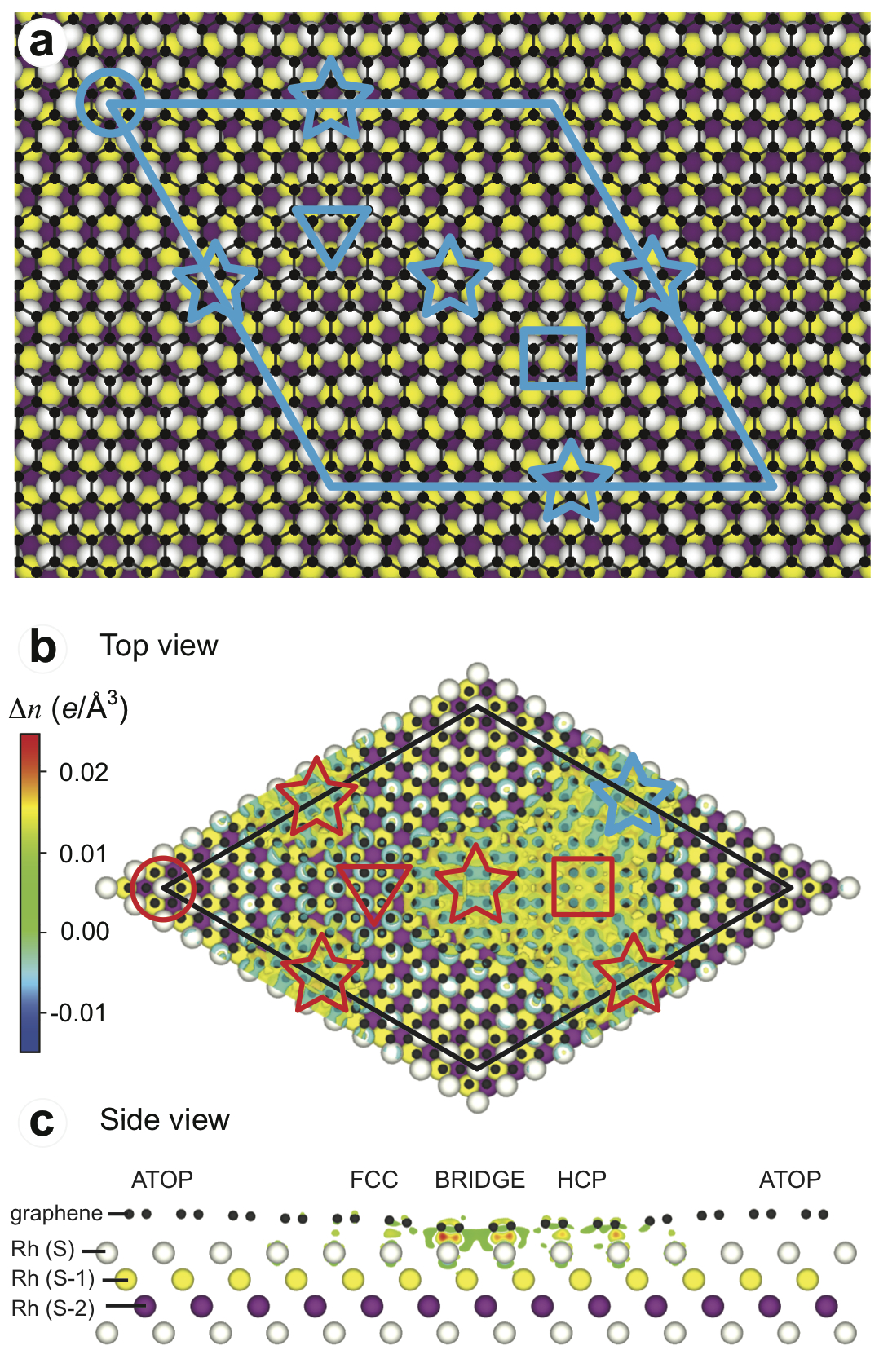}
\end{figure}

The graphene/Rh(111) can be considered as a representative example of the lattice-mismatched graphene-metal interfaces [Fig.~\ref{grRh_struct}(a)]. In this structure a graphene layer with $(12\times12)$ periodicity is arranged on $(11\times11)$ Rh(111) slab. Several high-symmetry adsorption sites for carbon atoms in a graphene layer can be identified in this structure depending on the stacking of underlying Rh: $ATOP$ [$A$; carbon atoms are placed above Rh(S-1) and Rh(S-2) atoms], $HCP$ [$H$; carbon atoms are placed above Rh(S) and Rh(S-2) atoms], $FCC$ [$F$;
carbon atoms are placed above Rh(S) and Rh(S-1) atoms], and $BRIDGE$ [$B$; Rh(S) atoms bridge the carbon atoms]. They are marked in Fig.~\ref{grRh_struct}(a) as circles, triangles, squares, and stars, respectively. Obviously that the local symmetry in the graphene/Rh(111) structure will define the strength of interaction between graphene and Rh(111) and among all high-symmetry positions the $BRIDGE$ places are expected to be the most energetically favourable for the nucleation of the adsorbed atoms and molecules.

The model shown in Fig~\ref{grRh_struct}(a) was used in the DFT-D2 structure-optimization procedure in order to get an information about the spatially distributed interaction strength in the graphene/Rh(111) system. The results of this step are summarised in Fig.~\ref{grRh_struct}(b,c) where the top and side views of the optimised structure of graphene/Rh(111) together with the distribution of the difference electron density,  $\Delta n(r)$, are shown. Carbon atoms in the $ATOP$ positions define the largest distance between graphene and Rh(111) of $3.15$\,\AA. The lowest position in this structure is the $BRIDGE$ one with the distance between graphene and Rh(111) of $2.08$\,\AA\ giving a maximum corrugation of the grapehene layer of $1.07$\,\AA. The $HCP$ and $FCC$ high-symmetry positions are lying by $0.4$\,\AA\ and $0.8$\,\AA, respectively, higher than the minima.

The obtained results point out the importance of the dispersive (van der Waals) forces for the description of the strongly-corrugated lattice mismatched graphene/metal interfaces. The test calculations performed without inclusion of the van der Waals interactions (standard DFT-PBE approach) give the larger corrugation of $1.8$\,\AA\ with a similar graphene/Rh distance for the $BRIDGE$ positions. At the same time the highest position for carbon atoms is lying at $3.90$\,\AA. This is due to the fact that standard GGA calculations give a reasonable result when describing the \textit{strong} interaction between graphene and metal, whereas the dispersion forces, neglected by the standard procedure, are important for the description of the \textit{weakly} interacting graphene/metal interfaces. The latter fact was shown to be valid when describing the \textit{weakly} adsorbed graphene layer on Ir(111), where bonding was explained by the existence of attractive van der Waals forces between graphene and Ir that modulated by weak bonding interactions at the $FCC$ and $HCP$ places and anti-bonding interaction around $ATOP$ positions~\cite{Busse:2011}. 

\begin{figure}
\caption{Carbon atom-projected total density of states ($\sigma$ and $\pi$) in the valence band for the different high-symmetry positions of the graphene/Rh(111) system. The inset shows the corresponding density of states for the $p_z$ character only.}
\label{dos}
\includegraphics[width=13cm,keepaspectratio]{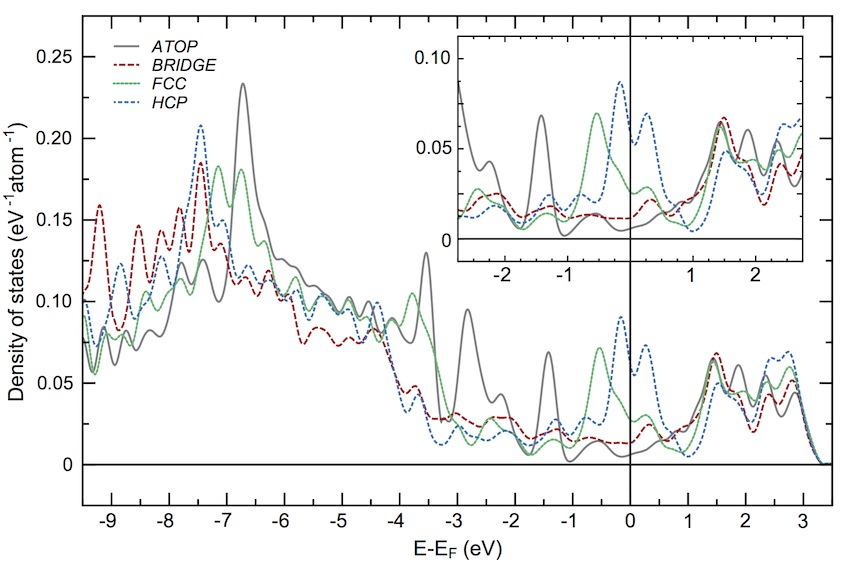}
\end{figure}

Figs.~\ref{grRh_struct}(b,c) show the top and side views of the graphene/Rh(111) interface with the difference electron density map,  $\Delta n(r)$: red colour - electron charge accumulation and blue colour - electron deficiency. One can clearly see that around the strongly bonded regions the electron density transfer from metal to graphene is observed: Rh $5sp$ electrons are more mobile compared to $4d$ and the transfer of these electrons on the graphene-derived $\pi^*$ states define the doping level of whole graphene which is in metallic state in this system. This effect is also reflected in the shift of the corresponding carbon-atom-projected density of states in the electronic structure of graphene/Rh(111): the graphene-derived states are shifted by $\approx1.5$\,eV to the larger binding energies compared to the free-standing graphene (Fig.~\ref{dos}). Additionally the sublattice symmetry for two carbon atoms in the graphene layer is broken at all places. However, only around the strongly-interacting regions it is accompanied with the strong hybridisation of the graphene $\pi$ states and Rh $4d$ states that leads to the opening of the energy gap for the $\pi$ states around the $K$ point of the Brillouin zone of graphene. This effect of hybridisation leads to the appearing of the so-called interface states which can be recognised in the DOS picture as a number of peaks around $E_F$ (Fig.~\ref{dos}). The similar effects were also observed for the \textit{strongly} interacting graphene/Ni(111) and graphene/Fe/Ni(111) systems~\cite{Bertoni:2004,Dedkov:2010a,Weser:2011,Voloshina:2011NJP}. 

\begin{figure}
\caption{(a) STM image ($150\times150$\,nm$^2$) of the clean Rh(111) surface. The inset ($1.9\times1.9$\,nm$^2$) shows the atomically resolved zoom of (a). (b-d) Carbon layers prepared on Rh(111) at different cracking temperatures: (b) 600\,K, image size $100\times100$\,nm$^2$, inset size $8\times8$\,nm$^2$; (c) 700\,K, image size $70\times70$\,nm$^2$, inset size $2.8\times2.8$\,nm$^2$ of region A; (d) 800\,K, image size $80\times80$\,nm$^2$, inset size $10\times10$\,nm$^2$ of the border region between regions C and D.}
\label{stm_overview}
\includegraphics[width=11cm,keepaspectratio]{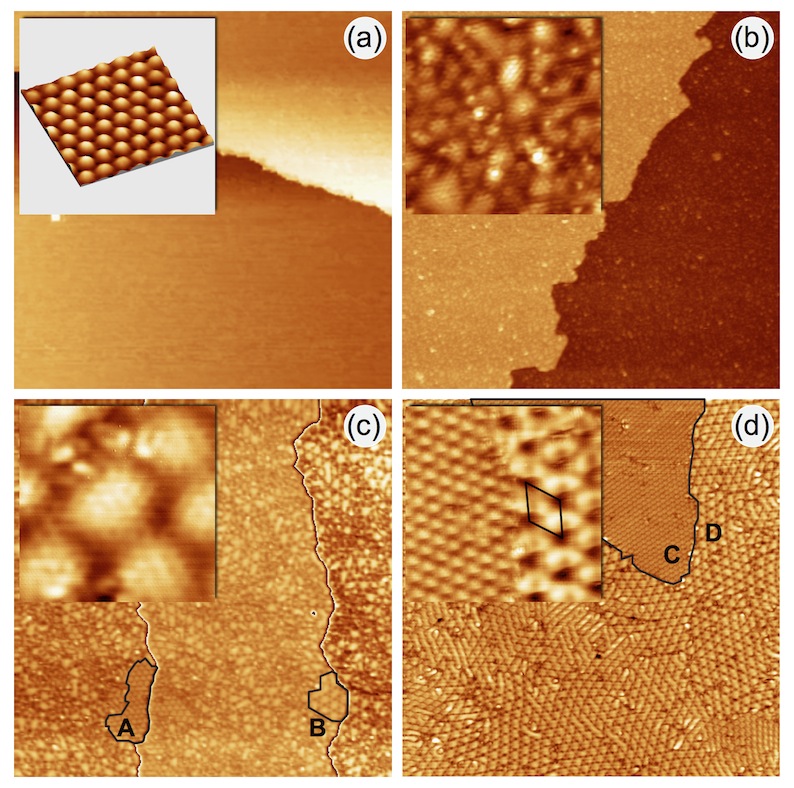}
\end{figure}

We have studied the temperature dependence of the graphene growth at different temperatures and the same pressure of the propylene gas ($p=2\times10^{-7}$\,mbar). The results are compiled in Fig.~\ref{stm_overview}. The growth of the ordered carbon layer (graphene) undergoes several stages where disordered carbon layer [Fig.~\ref{stm_overview}(b)] or mixture of several carbon phases (carbidic and graphene) are observed [Fig.~\ref{stm_overview}(c,d)]. Initially, the low-temperature cracking of propylene ($T=600$\,K) on the clean Rh(111) surface [Fig.~\ref{stm_overview}(a)] leads to the formation of the carbon layer without any long-range order [randomly distributed carbon-bubbles in Fig.~\ref{stm_overview}(b)] and only short-range order of carbon rings is observed [inset of Fig.~\ref{stm_overview}(b)]. Increasing of the synthesis temperature [$T=700$\,K and $T=800$\,K for Fig.~\ref{stm_overview}(c) and (d), respectively] leads to the ordering of the carbon layer and two carbon phases can be clearly distinguished [A (C) and B (D) on the corresponding STM images]. We assign the first phase [A in (c) and C in (d)] to the carbidic phase of carbon on Rh(111) and the second phase [B in (c) and D in (d)] to the graphene phase on Rh(111). Note: the studies of the process of the carbon phases on Rh(111) is out of the scope of the present manuscript. However, we would like to emphasise that the assignment of the later phases to the ordered graphene layer on Rh(111) is made on the facts that this phase starts to prevail at higher synthesis temperature and the observed STM images of this phase are in very good agreement with the calculated ones (see discussion below).

\begin{figure}
\caption{(a) Large scale STM image ($150\times75$\,nm$^2$) of the single domain graphene layer on Rh(111). The inset shows the corresponding LEED image. (b) Atomically resolved STM image ($6\times6$\,nm$^2$) of the moir\'e structure of the graphene/Rh(111) system. Tunnelling conditions: $U_T=-0.55$\,V, $I_T=10$\,nA. (c) Calculated STM image of graphene/Rh(111): integration was performed in the energy range $E-E_F=-0.5$\,eV and the distance between tip and the sample was set to 2\,\AA.}
\label{grRh_STM}
\includegraphics[width=11cm,keepaspectratio]{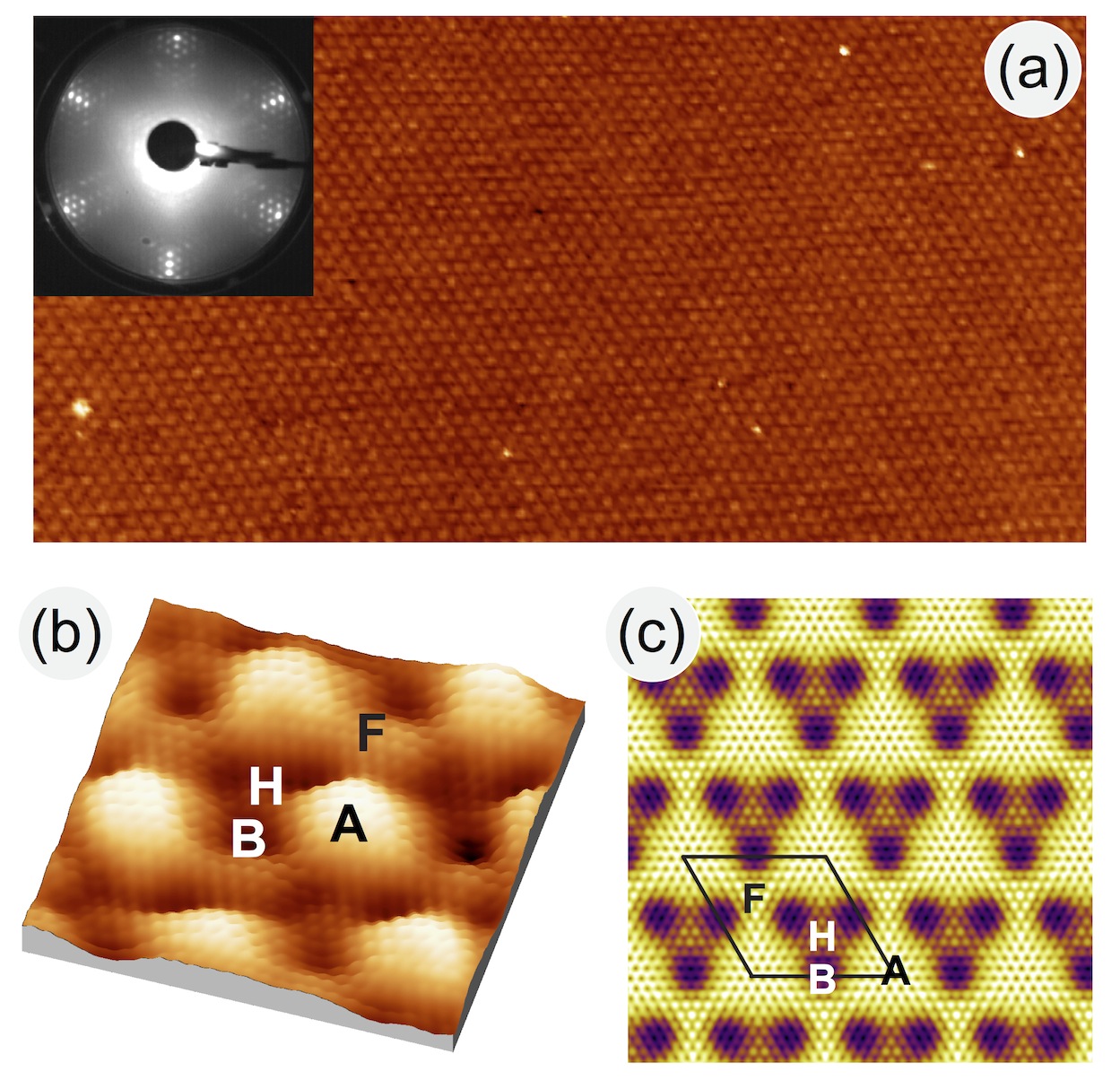}
\end{figure}

The further increasing of the cracking temperature to $T=1100$\,K results in the formation of the high quality single domain graphene layer on Rh(111) [Fig.~\ref{grRh_STM}(a)]. The corresponding LEED image of the obtained sample is shown as an inset. The zoomed STM image of the graphene moir\'e structure on Rh(111) is shown in Fig.~\ref{grRh_STM}(b) where one can clearly identify all high symmetry positions discussed earlier (these regions are marked by the respective capital letters). The obtained results are well reproduced in the theoretically calculated STM image of graphene/Rh(111) which is presented in Fig.~\ref{grRh_STM}(c): the corrugation as well as the difference in hight for different high-symmetry places are well confirmed. These results compiled on this figure are in very good agreement with earlier published data and the detailed discussion and interpretation of STM data can be found elsewhere~\cite{Wang:2010ky,Sicot:2010,Voloshina:2012a}.

The variation of the charge distribution [Fig.~\ref{grRh_struct}(b,c)] and the corresponding local density of states [Fig.~\ref{dos}] in the graphene moir\'e on Rh(111) gives a hint on the expectation of the most energetically favourable places of this structure for the adsorption and nucleation of the deposited atoms and molecules. With this respect the $HCP$, $FCC$, and $BRIDGE$ positions are the most probable places due to the increased local density of states at $E_F$ for these places as well as to the localisation of the negative charge around them. That can lead to the formation of the bonds (physisorption and chemisorption) between graphene and deposited atoms. In order to prove this idea we performed the atomic force microscopy/spectroscopy experiments on graphene/Rh(111). Here the oscillating tip can mimic the deposited metallic cluster and the interaction force between graphene and tip is reflected in the change of the resonance frequency of the sensor.

\begin{figure}
\caption{(Upper row, from left to right) Frequency shift plots ($\Delta f$), corresponding interaction force ($F$), and the tunnelling current between tip and sample as functions of the relative distance with respect to the scanning position ($U_T=-840$\,meV, $I_T=0.48$\,nA) for $ATOP$ ($A$) and $BRIDGE$ ($B$) positions of the graphene/Rh(111) system. (a-d) Constant frequency shift AFM images of graphene/Rh(111) measured at the frequency shifts of the sensor marked by the dashed lines and the corresponding letters in the figure for $\Delta f$ in the upper row.}
\label{z-sp_AFM}
\includegraphics[width=12cm,keepaspectratio]{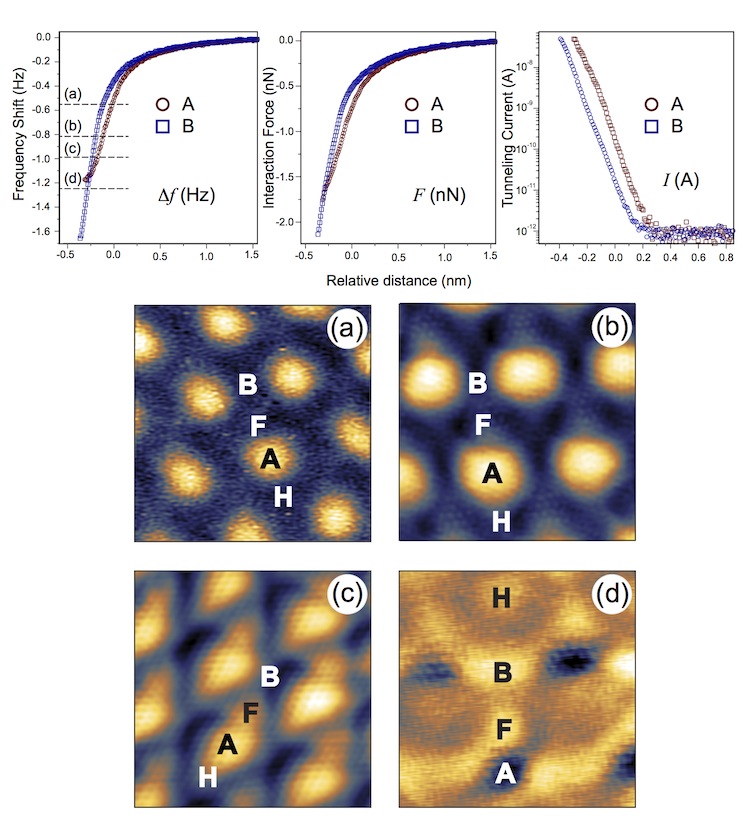}
\end{figure}

\begin{figure}
\caption{(a) Experimental constant frequency shift (CFS) AFM image of graphene/Rh(111). (b) and (c) Calculated CFS and constant hight (CH) AFM images of graphene/Rh(111).} 
\label{grRhAFM_detail}
\includegraphics[width=12cm,keepaspectratio]{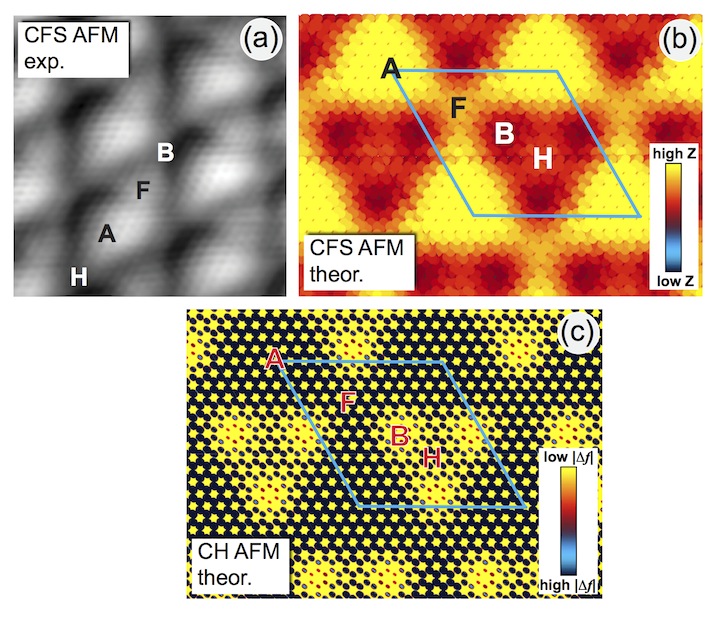}
\end{figure}

The results of these experiments and modelings are compiled in Figs.~\ref{z-sp_AFM} and ~\ref{grRhAFM_detail}. The resonance frequency shift of the oscillating scanning tip as well as the calculated force and the simultaneously measured tunnelling current as a function of the distance between tip and sample are shown in the upper row of Fig.~\ref{z-sp_AFM}. The \textit{zero}-level for the distance is taken with respect to the scanning position of the tip when the STM feedback loop is switched on to keep tunnelling current constant. One can clearly see that the strength of the interaction of tip and sample is different for for $ATOP$ (curve $A$) and $BRIDGE$ (curve $B$) positions as can be concluded from the discussion placed earlier. At the same time the tunnelling current measured in the same distance-dependent experiment shows the exponential dependence as expected.

Figs.~\ref{z-sp_AFM}(a-d) show the constant-frequency AFM images of graphene/Rh(111) acquired at set points marked by the dashed lines and the corresponding letters in the $\Delta f$ plots from the upper row. The respective high-symmetry places for carbon atoms of the graphene/Rh(111) structure are marked by the capital letters. One can clearly see that increasing of the frequency shift set-point during scanning leads to the increasing of the imaging contrast. Initially, the contrast between $hills$ and $valleys$ can only be distinguished. The further increasing of the frequency shift helps to resolve not only moir\'e structure of graphene/Rh(111), but also to get the atomic contrast in AFM images. This can be explained as due to the large contribution of the chemical forces (attractive contribution) when the frequency shift is increased. It is interesting to note that the constant frequency shift AFM imaging contrast for the graphene/Rh(111) is exchanged when one going from Figs.~\ref{z-sp_AFM}(c) to (d). This effect can be understood on the basis of the $\Delta f$ curves (upper row of Figs.~\ref{z-sp_AFM}) which intersect at $\Delta f\approx -1.15Hz$ indicating different $z$ set points when measuring at the frequency shifts corresponding to Figs.~\ref{z-sp_AFM}(c) and (d).

We also perform simulations of constant frequency shift and constant hight AFM images of graphene/Rh(111) (see discussion of the simulation details in the experimental part and in Ref.~\cite{Chan:2009jb}). Generally, the model analysis of the corrugated graphene-based systems was performed in Ref.~\cite{Castanie:2012bv} where the Lennard-Jones model potential between tip and sample was taken instead of the real interaction potential. Two contributions in the AFM imaging were considered: (i) due to the corrugation of the graphene-based system [in our case corrugation of graphene on Rh(111), i.\,e. geometrical effect] that shifts two interaction curves in $z$ direction and (ii) different strength of interaction between tip and sample for two different places as an effect of different interaction of graphene with a substrate at different places [in our case that can be connected with the different local density of states in graphene on Rh(111), i.\,e. electronic effect] that shifts two interaction curves in vertical direction.

The results of these simulations are presented in Fig.~\ref{grRhAFM_detail} where they are compared with the atomically resolved constant frequency shift AFM image of graphene/Rh(111). One can clearly see very good agreement between two AFM images, experimental and theoretical [Fig.~\ref{grRhAFM_detail}(a,b)] where the correct corrugation as well as details of AFM images are well reproduced. For the constant hight AFM images it is interesting to see that the imaging contrast for this structure is fully reversed with respect to constant frequency shift AFM image. This can be also understood on the basis of consideration of the attractive parts of the $\Delta f$ curves for different places of the graphene/Rh(111) structure (Figs.~\ref{z-sp_AFM}): The $ATOP$ ($A$) positions will give more negative frequency shift compared to the $BRIDGE$ ($B$) positions (in terms of the absolute values the $\Delta f$ value is larger for $A$ position). In case of the repulsive regime one can expect the opposite situation and the imaging contrast is reversed with respect to the attractive imaging regime. Unfortunately, the used approach can not give an information about absolute values of the frequency shift and the corrugation of the system (see Ref.~\cite{Chan:2009jb}). Moreover, the information about repulsive regime in AFM can not also be obtained from these simulations. However, if the model repulsive potential is added then the full simulation of the system can be performed.  

\section{Conclusion}

The geometry and electronic structure of the graphene/Rh(111) system were studied via combination of DFT, STM, and NC-AFM methods. We perform simulation of STM and AFM images on the basis of the optimised structure of graphene/Rh(111) and the charge distribution in this system and found that in both methods the imaging contrast is mainly defined by the structural corrugation of the system. The simulation of the imaging contrast in AFM was performed for the constant frequency shift and the constant hight methods and we found that it is inverted with respect to each other when forces between tip and sample are attractive. The obtained results shed light on the interactions in the graphene/Rh(111) system and help us to understand the observed effects in STM and NC-AFM imaging of the corrugated graphene-based systems.

\section*{Acknowledgements}

E.V. and E.F. acknowledge support from the DFG through the Collaborative Research Center (SFB) 765 and computing facilities (ZEDAT) of the Freie Universit\"at Berlin and of the North-German Supercomputing Alliance (HLRN).


\end{document}